# A Multi-wavelength Study of Parent Volatile Abundances in Comet C/2006 M4 (SWAN)

Michael A. DiSanti,[1] Geronimo L. Villanueva,[1,2] Stefanie N. Milam,[3] Lindsay N. Zack,[4] Boncho P. Bonev,[1,2] Michael J. Mumma,[1] Lucy M. Ziurys,[4] William M. Anderson[1,2]

[1] Solar System Exploration Division, NASA Goddard Space Flight Center, Greenbelt, MD 20771; michael.a.disanti@nasa.gov
[2] Department of Physics, The Catholic University of America, Washington, DC 20064
[3] SETI Institute and NASA-Ames Research Center, Moffett Field, CA 94035
[4] Department of Chemistry, Department of Astronomy, University of Arizona, Tucson, AZ 85721

**Abstract**

Volatile organic emissions were detected post-perihelion in the long period comet C/2006 M4 (SWAN) in October and November 2006. Our study combines target-of-opportunity observations using the infrared Cryogenic Echelle Spectrometer (CSHELL) at the NASA-IRTF 3-m telescope, and millimeter wavelength observations using the Arizona Radio Observatory (ARO) 12-m telescope. Five parent volatiles were measured with CSHELL ($H_2O$, $CO$, $CH_3OH$, $CH_4$, and $C_2H_6$), and two additional species (HCN and CS) were measured with the ARO 12-m. These revealed highly depleted CO and somewhat enriched $CH_3OH$ compared with abundances observed in the dominant group of long-period (Oort cloud) comets in our sample and similar to those observed recently in Comet 8P/Tuttle. This may indicate highly efficient H-atom addition to CO at very low temperature (~ 10 – 20 K) on the surfaces of interstellar (pre-cometary) grains. Comet C/2006 M4 had nearly "normal" $C_2H_6$ and $CH_4$, suggesting a processing history similar to that experienced by the dominant group. When compared with estimated water production at the time of the millimeter observations, HCN was slightly depleted compared with the normal abundance in comets based on IR observations but was consistent with the majority of values from the millimeter. The ratio CS/HCN in C/2006 M4 was within the range measured in ten comets at millimeter wavelengths. The higher apparent H-atom conversion efficiency compared with most comets may indicate that the icy grains incorporated into C/2006 M4 were exposed to higher H-atom densities, or alternatively to similar densities but for a longer period of time.

Keywords: Comets, composition; Comets, origin; Molecular spectroscopy; Comet C/2006 M4 (SWAN)

1. INTRODUCTION

Comets formed relatively far from the Sun, beyond the "frost line," and they reside for long periods of time in the outer Solar System. Their abundances can therefore provide clues to the formation and evolution of the Solar System. The original structures and compositions of ices contained within their nuclei (i.e., native ices) should reflect local conditions (chemistry, temperature, degree of radiation processing) prevalent when and where they formed (e.g., DiSanti and Mumma 2008, Bockelée-Morvan et al. 2004, Irvine et al. 2000, Mumma et al. 1993). Although comets contain relatively primitive icy material remaining from the epoch of Solar System formation, the extent to which they are modified from their initial state is a fundamental question in cometary science.

The current orbits of comets provide information on their recent dynamical history, however tracing cometary origins is complicated by their radial migration in the proto-planetary disk and by dynamical interactions with the growing giant planets (Levison and Morbidelli 2003, Gomes et al. 2005[5], Tsinganis et al. 2005[5], Morbidelli et al. 2008). Such interactions placed comets into (at least) two principal present-day reservoirs: the Oort cloud (OC) and the Edgeworth-Kuiper belt (KB), thought to be the principal source regions for "nearly-isotropic" (long-period, dynamically-new, and Halley-type) comets and "ecliptic" (e.g., Jupiter-family) comets (JFCs), respectively. Comet nuclei remain largely unaltered until perturbed into the inner Solar System (Stern 2003, Gladman 2005), at which time sublimation of surface (or near-surface) native ices releases parent volatiles into the coma where they can be measured spectroscopically.

Spectroscopic investigations reveal chemical diversity among comets, based on studies of parent volatiles at infrared (e.g., Mumma et al. 2003, DiSanti and Mumma 2008) and radio wavelengths (e.g., Biver et al. 2002), implying a range of natal conditions experienced by pre-cometary ices (see also Bockelée-Morvan et al. 2004). Fundamental clues to natal conditions can be provided both by measuring native ice abundances within an individual comet, and by building a taxonomy through compositional comparisons among comets from both OC and KB populations.

This paper combines post-perihelion infrared and millimeter spectral observations of the long-period comet C/2006 M4 SWAN (hereafter C/2006 M4) that characterize its native ice composition. Discovered on 20 June 2006 with the Solar Wind ANisotropy instrument (SWAN) of the Solar Heliospheric Observatory (Matson and Mattiazzo 2006 IAUC 8729), C/2006 M4 was initially assigned an eccentricity $e = 1.0$ based on 39 observations spanning 12-21 July 2006 (Teamo and Hoenig 2006 IAUC 8733; see also MPEC 2006-N38).

---

[5] In particular, these papers detail the recently developed "Nice Model."

Inclusion of additional astrometric observations led to a slightly hyperbolic orbit (see JPL's HORIZONS website: url http://ssd.jpl.nasa.gov/horizons.cgi), suggesting that C/2006 M4 may be entering the solar system for the first time (i.e., that it may be dynamically new). However, the original value for 1/a (0.000194 AU$^{-1}$; Nakano 2008) places its aphelion at $1.03 \times 10^4$ AU, indicating a dynamical origin in the inner Oort cloud reservoir and calling into question its dynamical status (returning long-period versus new). In any case, our observations occurred post-perihelion, so even if dynamically new any radiation (or cosmic ray) processed outer layers (over its residence time in the Oort cloud) would likely have been lost through erosion prior to our observations, making our measurements a probe of less processed underlying material.

Visual estimates projected C/2006 M4 to reach magnitude 6 (or brighter) by mid October 2006 (IAUC 8766; CBET 738), however it experienced approximately a 1.5-2-magnitude surge in brightness around 24 October, followed by a sharp decline (Yoshida 2007); we incorporate this into our analysis in Section 4.2. Optical imaging revealed a prominent ion tail but no discernable dust tail, suggesting that the observed coma was composed primarily of gaseous species. This made C/2006 M4 a good target for our study.

## 2. OBSERVATIONS

2.1 *Target-of-opportunity observations at the NASA Infrared Telescope Facility*

The comet was discovered after the normal application deadline for the Fall 2006 proposal cycle. A target-of-opportunity request was submitted to use the long slit Cryogenic Echelle Spectrometer (CSHELL; Tokunaga et al. 1990) at the NASA Infrared Telescope Facility on Mauna Kea, HI, based on its favorable geocentric velocity (see Table 1) and (especially) its availability during daylight hours since night time observing was fully scheduled.

A total of eleven hours of clock time were granted at the IRTF. We used CSHELL with a 1 arc-second-wide slit, resulting in a spectral resolving power ($\lambda/\Delta\lambda$) of ~ 25,000 that permitted studies of individual line intensities. Flux calibration was achieved through observations of the IR standard star BS7235 using a wide (4 arc-second) slit.

The CSHELL observations targeted specific emissions from CO, $CH_4$, $C_2H_6$, and $CH_3OH$, and these were measured simultaneously or nearly simultaneously with $H_2O$ (or its proxy OH*) to ensure reliable mixing ratios between trace constituents and water (Fig. 1). The significant geocentric velocity (Δ-dot) of C/2006 M4 displaced its lines of $CH_4$ and CO from their opaque telluric counterparts to regions of exceptionally high atmospheric transmittance (80 – 90 percent).

[INSERT FIGURE 1 HERE]

Measuring HCN in comets represents a principal means of probing nitrogen chemistry in the early solar system, and is fairly routine with modern infrared spectrometers, even in less productive comets (see, e.g., Magee-Sauer et al. 1999, Magee-Sauer et al. 2002, Mumma et al. 2003, Bockelée-Morvan et al. 2004, Magee-Sauer et al. 2008, DiSanti and Mumma 2008, and references therein). Our observing plan included a CSHELL setting that targets both HCN and $C_2H_2$. However, we were unable to utilize it due to the extremely limited amount of observing time available for C/2006 M4, especially considering that a stringent test of abundances (particularly that of $C_2H_2$) would have required a time on source comparable to that spent on the setting that samples $H_2O$ emission near 3450 $cm^{-1}$ (Table 1).

We used the standard ABBA observing sequence, with an east-west (along-slit) nod of 15 arc-seconds (half the slit length) between A- and B-beam positions (e.g., see discussions in Dello Russo et al. 2004, DiSanti et al. 2006). The observations were conducted during daylight, so the CSHELL CCD guider could not be used. Instead, following every 1 – 2 ABBA sequences (approximately 5 – 10 minutes of clock time) the comet was imaged through the CSHELL open (30 x 30 arc-second) aperture, to re-position it in the slit and to update tracking rates as warranted. Cometary drift was generally 1 – 2 arc-seconds or less (approximately commensurate with the seeing conditions), and was primarily along the length of the slit. This preserved cometary signal in the slit and made it possible (upon reduction), based on temporal interpolation of comet positions in the images, to spatially register all A and B scans separately prior to combining them.

The low dust continuum emission made imaging the comet during daytime challenging. We found that its visibility peaked near 3.52 $\mu$m, corresponding to the approximate center of the $CH_3OH$ $\nu_3$ band (with Q-branch at 3.516 $\mu$m, or 2844 $cm^{-1}$), as methanol was relatively abundant in C/2006 M4 (see below) and so contributed most of the intensity at this wavelength. To minimize potential shifts in wavelength calibration and mismatches in flat fielding, the echelle grating was kept stationary for the duration of each given setting. The CSHELL low resolution ($\lambda/\Delta\lambda \sim 80$) circular variable filter (used for transmitting only the desired echelle order) was tuned to 2844 $cm^{-1}$ for imaging the comet between spectral sequences.

[INSERT TABLE 1]

2.2 *Millimeter observations with ARO 12-m telescope*

Observations of C/2006 M4 were conducted using the Arizona Radio Observatory (ARO) 12-m telescope on Kitt Peak, AZ (Table 1). The ARO 12-m observations targeting HCN and CS in C/2006 M4 were conducted during several observing runs spanning 2006 October 30 to November 10. Dual-channel SIS mixers operated in single-sideband mode with image rejection around 20 dB were employed in the 2 mm (for the J = 3 – 2 transition of CS) and 3 mm (for J = 1 – 0 of HCN) bands, and filter banks having resolutions of 250 and 100 kHz, respectively, were used as back-ends. The spectral temperature scale was determined by the chopper-wheel method, corrected for forward spillover losses, and is given in terms of antenna temperature, $T_R^*$ (K). The radiation (or equivalent brightness) temperature, $T_R$, is then derived from the corrected beam efficiency, $\eta_c$, as $T_R = T_R^*/\eta_c$. The ephemeris employed was JPL reference orbits #10 for October and #11 for November. Nearby planets were used to check focus and positional accuracy at regular intervals. Data were taken in position-switching mode with the off position located +30 arc-minutes west in azimuth. HCN and CS were clearly detected in the millimeter observations (Fig. 2).

At millimeter wavelengths, HCN is the standard against which abundances of other cometary volatiles (both parent and daughter species) are measured, and the limited observing time available for C/2006 M4 with CSHELL made the millimeter observations of paramount importance. CS tests the role of sulfur in the formative region and provides an additional comparative among comets (see discussion in Section 4.4, and Biver et al. 2002, Biver et al. 2006).

[INSERT FIGURE 2]

**3. DETERMINATION OF MOLECULAR ABUNDANCES**

The production rate (Q, molecules s$^{-1}$) of a parent volatile is expressible for both IR and millimeter observations in terms of the column density $N$ (molecules m$^{-2}$) in the beam:

$$Q = \frac{N \, A_{beam}}{\tau_{1AU} f(x)} R_h^{-2} \, q_{scale}. \quad (1)$$

Here $A_{beam}$ (m$^2$) is the area subtended by the beam, $\tau_{1\,AU}$ (s) is the photo-dissociation lifetime of the parent molecule evaluated at heliocentric distance $R_h$ = 1 AU, and f(x) represents the fraction of all molecules in the coma that are encompassed within the beam. The dimensionless factor $q_{scale}$ accounts for loss of flux due to aperture effects or beam dilution.

### 3.1 *Production rates from infrared spectra*

We used customized IR data processing algorithms to produce orthogonal spatial-spectral frames from the raw CSHELL data (Villanueva et al. 2006, DiSanti et al. 2006, DiSanti et al. 2001). Wavelength calibration was applied to these processed data through comparison of observed sky radiance with spectra synthesized using a rigorous line-by-line, layer-by-layer radiative transfer model of the terrestrial atmosphere (GENLN2, Edwards 1992). We updated this model to properly include pressure-shift coefficients along with the latest spectroscopic parameters (Villanueva et al. 2008).

Assuming optically thin conditions, column densities were determined from the CSHELL data using line fluxes ($F_{line}$, W m$^{-2}$), corrected for atmospheric transmittance at each Doppler-shifted line frequency, and fluorescence g-factors (W molecule$^{-1}$):

$$N_{IR} = \frac{4\pi}{\Omega} \frac{F_{line}}{g_{line,1AU}} R_h^2, \qquad (2)$$

in which $\Omega$ (ster) is the solid angle subtended by the beam – we used 15-row spectral extracts (i.e., a 1x3 arc-second aperture) centered on the nucleus for our analysis. This translated to approximately 800x2400 km at the comet on UT November 7, and 840x2500 km on November 10. For the CSHELL observations, the fraction f(x) in Eq. 1 is calculated for square pixels (with x being the ratio of pixel size to photo-dissociation scale length; see the Appendix of Hoban et al. 1991 for details) assuming constant gas production and uniform outflow at constant speed (taken to be 800 $R_h^{-0.5}$ m s$^{-1}$, consistent with line widths measured in the ARO spectra; see below).

The factor $q_{scale}$ (> 1 for our CSHELL observations) corrects for slit losses due to seeing and to potential drift of the comet arising from uncorrected errors in telescope tracking, in particular as arise from (slight) drift perpendicular to the slit. The production rate corresponding to $q_{scale}$ = 1.0 is referred to as the nucleus-centered Q; $q_{scale}$ is typically determined from the sum of spatial profiles for several lines because this summed profile has higher signal-to-noise compared with individual line profiles. Applying $q_{scale}$ to the nucleus-centered Q for each line yields its global Q. The overall molecular production rate is then the weighted mean of line-by-line global Qs (Table 2). The methodology for determining molecular production rates is detailed in the literature (e.g., Magee-Sauer et al. 1999, Dello Russo et al. 2000, DiSanti et al. 2001, DiSanti et al. 2006, Magee-Sauer et al. 2002, Bonev 2005).

[INSERT TABLE 2]

### 3.2 *Production rates from millimeter observations*

Abundances for HCN and CS were derived from the ARO 12-m observations assuming the source filled the (circular) beam of the telescope. Again assuming the emissions to be optically thin (based on intensities of HCN hyperfine components; see Fig. 2A), the column density was calculated as:

$$N_{radio} = \frac{3k \int T_R dv_{1/2}\, \zeta_{rot}}{8\pi^3 \nu\, S_{ij}\mu_o^2\, \mathbf{g}_j\, e^{-E_j/kT_{rot}}}, \tag{3}$$

in which $\nu$ (MHz) is the frequency, $\int T_R dv_{1/2}$ (K km s$^{-1}$) represents the integrated intensity corrected for main beam efficiency, $S_{ij}\mu_o^2$ is the line strength ($\mu_o$ being the permanent dipole moment, in Debye), $\zeta_{rot}$ is the rotational partition function, and $\mathbf{g}_j$ and $E_j$ are respectively the statistical weight and energy of the upper rotational state.

For the millimeter observations, the factor $A_{beam}$ from Eq. 1 is given by $\pi(D_{beam}/2)^2$, where the beam diameter $D_{beam} = \Delta \tan(1.22\lambda/D_{tel})$ and $D_{tel}$ = 12 m; this translates to approximately $D_{beam}$ = 5.44 x 10$^4$ km at the comet for the HCN observations and 3.28 x 10$^4$ km for the CS observations. Assuming the source fills the beam, $q_{scale}$ = 1.0 and, if solely native production and spherically-symmetric gas outflow at constant speed in the coma are assumed, f(x) can be calculated following Yamamoto (1982). Using a Monte Carlo formalism (as was done here) results in essentially the same production rates as those given by the strictly spherically-symmetric case. Terrestrial atmospheric opacity is included in the calibration of the millimeter observations.

HCN and CS were assumed to be parent species, and their production rates were determined using a Monte Carlo model for a purely native source (Milam et al. 2006; Remijan et al. 2008). The model traces the trajectories of molecules that are ejected from the comet surface and enter the telescope beam. Input parameters include molecular lifetime, gas outflow velocity, telescope beam size, and observed column density. From our line profiles (Fig. 2) we measured outflow velocities of ~ 730 m s$^{-1}$ and ~ 1000 m s$^{-1}$ respectively for HCN and CS, and adopt 800 $R_h^{-0.5}$ m s$^{-1}$ in our analysis, as stated previously.

If CS is a photo-destruction product of CS$_2$ (Snyder et al. 2001), only slight adjustments to the input parameters are introduced in calculating Q. Under this assumption, lifetimes and velocities of *both* parent (CS$_2$) and daughter (CS) are used, while other input parameters are held unchanged from their values assuming CS itself is a parent species. Because CS$_2$ has a very short photo-dissociation scale length (~ 300 km, corresponding to $\tau$ ~ 370 $R_h^2$ s; Huebner et al. 1992), much smaller than the beam radius (1.64 x $10^4$ km), this translates to a negligible change in $Q_{CS}$ compared with its value assuming CS to be a parent.

[INSERT TABLE 3]

3.3 *Rotational temperatures*

For both the CSHELL and ARO observations, a small number of lines were measured. For this reason, knowledge of rotational temperature ($T_{rot}$) in the coma is needed to establish the population distribution among rotational levels and thereby determine accurate column densities from measured line intensities. (The dependence on $T_{rot}$ is explicitly manifested through $g_{line}$ in Eq. 2, and through $e^{-E_j/kT_{rot}} / \zeta_{rot}$ in Eq. 3.)

Obtaining a reliable measure of $T_{rot}$ requires measuring lines that sample a range in rotational energy. Because each CSHELL setting encompasses a limited spectral range (approximately 0.23% of the central wave number), and also due to the limited IRTF time available to measure the chemistry of C/2006 M4, only the 3452 $cm^{-1}$ water setting satisfied this criterion – we measured $T_{rot}$ = 77 $^{+29}/_{-14}$ K for $H_2O$ on November 10, based on four lines sampling rotational energies ranging from ~ 40 – 400 $cm^{-1}$ (see inset to Fig. 1E). We therefore adopt $T_{rot}$ = 80 K in reporting global production rates for molecules observed with CSHELL (Table 2), and list column densities and production rates for HCN and CS assuming 60, 80, and 100 K (Table 3). In Table 4, we report mixing ratios (i.e., abundances relative to $H_2O$) for 60, 80, and 100 K for species measured with CSHELL. For HCN and CS, we present mixing ratios for 60 and 80 K, incorporating differences in estimated overall gas production between ARO and CSHELL observations resulting from the reported outburst (see Section 4.2).

**4. DISCUSSION**

Our combined IR and millimeter results reveal an interesting parent volatile composition for C/2006 M4. Some intriguing possibilities for the processing history of its pre-cometary ices emerge, particularly when compared with other comets previously characterized. A comparison among comets provides a context for inferring where (perhaps also when, for how long, and/or to what degree) processing occurred.

*4.1. Dependence of abundances on rotational temperature*

In Table 4 we list mixing ratios (abundances relative to $H_2O$) at $T_{rot}$ = 60, 80, and 100 K for CO, $C_2H_6$, $CH_3OH$, and $CH_4$ in C/2006 M4. The abundances of CO, $C_2H_6$, and $CH_3OH$ are independent of assumed rotational temperature to well within $1\sigma$ uncertainty. This is primarily because the settings targeting their emissions include multiple lines that sample a range of rotational energies. The abundance of $CH_4$ is somewhat more sensitive to $T_{rot}$, as is that of HCN (Table 4).

[INSERT FIGURE 3]

Assuming $T_{rot}$ for the ARO observations agrees with that measured with CSHELL for $H_2O$, the abundance of HCN can be expressed relative to the weighted mean of $H_2O$ production rates for November 7, 9, and 10 (Fig. 3), however this comparison is complicated. The aperture for CSHELL (~ 800 x 2400 km; Section 3.1) is small compared with the ARO beam sizes (5.4 x $10^4$ km for HCN, 3.3 x $10^4$ km for CS; Section 3.2), so we must allow for possible differences in beam-averaged $T_{rot}$. Perhaps more significantly, the ratio $HCN/H_2O$ cannot be adequately assessed without also accounting for differences in modeled water production rate between times of ARO and CSHELL observations due to the decaying outburst (Table 4 note e, Fig. 4, and Section 4.2). The abundance of CS is expressed relative to HCN (see Section 4.4) as is customary with millimeter observations (Biver et al. 2002, Biver et al. 2006; see Section 4.4).

[INSERT TABLE 4]

*4.2. Accounting for temporal variability*

The temporal coverage of HCN and CS with ARO encompassed that of the CSHELL observations (Table 1), however three of the four dates on which HCN was observed and two of four dates on which CS was observed were prior to our CSHELL run. C/2006 M4 experienced a significant outburst on UT 24 October followed by a steep decline in activity (see Yoshida 2007). Therefore, we also report a mixing ratios for HCN and CS based on the estimated contemporaneous $H_2O$ production rate for each.

To accomplish this, we reproduced the light curve of C/2006 M4 using fitted expressions that relate its visual magnitude to $R_h$ and $\Delta$ (Yoshida 2007). We then applied the expression (Jorda et al. 2008) relating $Q(H_2O)$ to visual magnitude (reduced to $\Delta = 1$ AU) based on 234 measurements in 37 comets. The resulting "activity curve" is shown in Fig. 4 – we note that our points from Fig. 3 are grouped tightly about this fit. Based on the weighted mean UT date for the HCN observations (November 3.0), this suggests a commensurate fitted value of $Q(H_2O) = 2.05 \times 10^{29}$ molecules $s^{-1}$. We list our "most probable" HCN abundance (0.13 ± 0.02 percent), assuming $T_{rot}$ = 60 K for HCN and $T_{rot}$ = 80 K for $H_2O$. For comparison, we also list its abundance assuming both HCN and $H_2O$ are characterized by $T_{rot}$ = 80 K (see Table 4, note 'e'). Therefore, unless $T_{rot}$ for HCN is well above 100 K (which seems very unlikely), HCN is depleted in C/2006 M4 compared with its "normal" abundance as measured in the IR (see discussion in Section 4.4).

However, our most probable mixing ratio for HCN is consistent with values found for the majority of 24 comets observed at millimeter wavelengths (see Figs. 1 and 2 of Biver et al. 2002, and Section 4.4). This may reflect what appears to be differences between HCN abundances derived from IR and millimeter observations, the latter being systematically lower by a factor of ~ 2 among measured comets (e.g., see Magee-Sauer et al. 2008, and references therein). This can be tested through coordinated observations.

[INSERT FIGURE 4]

*4.3. Relation to the chemistry of oxidized carbon*

An important recognized path for converting CO to the chemically linked molecules $H_2CO$ and $CH_3OH$ is H-atom addition reactions on surfaces of icy interstellar grains prior to their incorporation into the nuclei of comets (Hudson and Moore 1999, Watanabe and Kouchi 2002, Hiraoka et al. 2002, Watanabe et al. 2004). This process is analogous to production of cometary ethane through hydrogenation of $C_2H_2$, proposed to explain the relatively high abundance of $C_2H_6$ to $CH_4$ first observed in C/1996 B2 (Hyakutake) (Mumma et al. 1996). Subsequent studies routinely reveal high $C_2H_6$ abundances in comets, orders of magnitude higher than the amount of $C_2H_6$ produced through gas phase nebular chemistry (Prinn and Fegley 1989). This demonstrates that H-atom addition is likely a common and important process in the evolution of such grains.

Laboratory experiments show hydrogenation of CO to be efficient only at very low temperatures (~ 10 – 20 K), the yields being highly dependent on fluence (i.e., density) of atomic hydrogen, temperature (i.e., H-atom retention time), and whether CO is housed structurally with $H_2O$ in the ice (e.g., see Watanabe et al. 2004). One measure of this conversion can be represented by the sum of $H_2CO$ and $CH_3OH$ abundances divided by the sum of all three abundances (DiSanti et al. 2002).

Due to the limited amount of available observing time, we were unable to observe $H_2CO$ in C/2006 M4. Despite this, the severely depleted abundance of CO coupled with somewhat enriched $CH_3OH$ (Table 4) suggests highly efficient conversion of CO, approximately 90 percent or higher -- note that, by the above definition, the presence of non-negligible $H_2CO$ would further increase the conversion efficiency. This measure assumes that all CO converts to $H_2CO$ or $CH_3OH$, and that these two molecules arise solely from H-atom addition to CO, thereby representing a limiting case.

This interpretation of measured abundances may be compromised if icy grains in the inner proto-solar nebula were thermally processed prior to their final incorporation into the nucleus. It is possible that the grains incorporated into C/2006 M4 originally had a higher endowment of CO that later was preferentially lost through vaporization in subsequent warmer nebular environs. In this case one might also expect depletion of the only slightly less volatile $CH_4$, however methane falls in the "normal" range in this comet (Table 4). Alternatively, some fraction of the initial CO budget may subsequently have been locked up in complex organic material, for example the "CHON" particles discovered during the Giotto encounter with comet 1P/Halley (Kissel et al. 1986; Huebner 1987; Mitchell et al. 1987; Milam et al. 2006). The very weak observed optical / IR continuum in C/2006 M4 implies a low abundance of approximately micron-sized grains, so a commensurate paucity of smaller grains (e.g., CHON) might be expected. In any case, observed and primordial (pre-cometary) abundances of CO could differ substantially, and our measured conversion efficiency may reflect processing history of ices more than natal conditions in the proto-solar cloud.

Inter-comparison of abundances for CO, $H_2CO$, and $CH_3OH$ in comets, molecular disks around young stars, and proto-stellar cloud cores would be particularly valuable for understanding the (potential) processing (chemical and/or thermal) experienced by organic matter during different stages of planetary system formation. However, such studies are beyond the scope of the present paper.

*4.4. Abundance comparisons with other comets*

Comparisons among comets suggest interesting possibilities. The molecules in Table 4 are listed from left to right in order of increasing vacuum sublimation temperature. We arbitrarily adopt the following classifications for mixing ratios relative to those measured in four OC comets (which we refer to tentatively as "organics-normal" comets; see notes 'a' and 'f' in Table 4). "Severely" refers to values that are three standard deviations (3σ) or more <u>and</u> more than a factor of two from normal, while "slightly" (or "somewhat") refers to values between 1σ and 3σ or within a factor of two from normal. A range of native ice compositions is seen even among the small number of comets measured in detail (about two dozen between IR and millimeter regimes; Mumma et al. 2003, Bockelée-Morvan et al. 2004, Crovisier 2007, DiSanti and Mumma 2008 and references therein)[6].

Organics-normal comets show chemical signatures similar to (although not identical to) those observed in dense interstellar clouds (Mumma et al. 2003; Bockelée-

---

[6] Note that these classifications are based on small number statistics.

Morvan et al. 2004; Charnley and Rodgers 2008), suggesting the ices incorporated into the nuclei of these comets experienced processing at low temperatures. Compared with the composition-normal group of comets, our measurements of C/2006 M4 indicate severely depleted CO and somewhat enriched $CH_3OH$, while $CH_4$ and $C_2H_6$ are consistent with normal. After accounting for estimated differences in gas production between IRTF and ARO observations (Fig. 4), our abundance of HCN is somewhat depleted relative to normal IR values in comets but is consistent with that found in the majority of comets measured at millimeter wavelengths (see Table 4, note 'e'), which cluster near $HCN/H_2O$ ~ 0.1 percent (see Figs. 1 and 2 in Biver et al. 2002). Our abundance ratio CS/HCN is 0.63±0.101 and 0.59±0.094 for $T_{rot}$ = 60 K and 80 K, respectively; this incorporates differences in the mean UT for HCN and CS observations assuming their production rates track that estimated for $H_2O$ (Fig. 4). The ratio CS/HCN in C/2006 M4 falls within the range (0.5 – 1.2) measured at millimeter wavelengths in nine of 24 comets (Biver et al. 2002) plus 153P/Ikeya-Zhang (Biver et al. 2006). Of these ten comets, seven have probable Oort cloud origin and three are JFCs.

The hyper-volatiles CO and $CH_4$ exhibit the largest variations in abundance among comets, however they are not correlated, demonstrating that thermal considerations alone cannot explain molecular abundances in comets (Gibb et al. 2003). Fully interpreting the large variation in CO abundances among comets involves distinguishing the amount initially condensed onto grains versus the degree of subsequent alteration, for example through H-atom addition or incorporation into complex molecules as discussed previously.

Inspection of Table 4 shows that our abundances of CO and $CH_3OH$ in C/2006 M4 were very similar to those in Comet 8P/Tuttle (Böhnhardt et al. 2008), the composition of which also suggested highly efficient conversion of $C_2H_2$ to $C_2H_6$ (exceeding 85%). (Because $C_2H_2$ was not measured in C/2006 M4, this comparison is not possible.) However, C/2006 M4 had normal abundances of $CH_4$ and $C_2H_6$, 2 – 3 times higher than their corresponding values in 8P/Tuttle (Bonev et al. 2008; Böhnhardt et al. 2008). Provided our "most probable" abundance for HCN in C/2006 M4 is valid, this makes HCN only somewhat less depleted than its value in Comet Tuttle (Bonev et al. 2008), notwithstanding the systematic difference between HCN abundances based on IR and millimeter observations mentioned previously (Section 4.2). We note that preliminary results from IRAM and CSO measurements (Biver et al. 2008a) place the abundance of HCN in 8P/Tuttle towards the low end of the distribution measured among comets.

Low CO abundances were also observed in long-period comet C/1999 S4 (LINEAR) and JFC 73P/Schwassmann-Wachmann 3, however in contrast to C/2006 M4 these comets showed severe depletions in most parent volatiles, including $CH_3OH$ but not HCN (Mumma et al. 2001, Villanueva et al. 2006, Kobayashi et al. 2007, Dello Russo et al. 2007, DiSanti et al. 2007a). Enriched abundances were observed in long-period comet C/2001 A2 (Biver et al. 2006, Gibb et al. 2007, Magee-Sauer et al. 2008), and more recently in JFC 17P/Holmes during its sudden outburst in October/November 2007 (Salyk et al. 2007, Dello Russo et al. 2008, Bockelée-Morvan 2008, Biver et al. 2008b), suggesting a more nearly interstellar chemistry. For both relatively depleted and enriched comets, this demonstrates that compositional similarities exist between comets that likely come from different dynamical reservoirs (OC for C/1999 S4 and C/2001 A2, KB for 73P and 17P), and suggests distinct processing histories for depleted, organics-normal, and enriched comets. Our results for C/2006 M4 appear to be most closely in line with the normal group of comets, although differences exist as discussed above.

If formation temperatures were comparable to those experienced by organics-normal comets, and assuming H-atom addition dominates conversion of CO (see Section 4.3), it is possible that the higher $CH_3OH/CO$ ratio in C/2006 M4 implies a greater availability (e.g., a higher density) of atomic hydrogen in its formation environment. Alternatively, H-atom addition reactions could have taken place over a longer period of time in the case of C/2006 M4, leading to excess methanol at the expense of CO.

**5. SUMMARY**

We conducted post-perihelion observations of parent volatiles in the long-period comet C/2006 M4 (SWAN) on three UT dates using CSHELL at the NASA-Infrared Telescope Facility 3-m telescope, and on seven UT dates using the Arizona Radio Observatory 12-m telescope. Our measured abundances for $C_2H_6$, $CH_4$, HCN, CO, $CH_3OH$, and CS were compared to their values in other comets, as part of our ongoing efforts to build a taxonomy based on composition. The values for $C_2H_6$ and $CH_4$ were consistent with those found in the organics-normal class of comets. After accounting for differences in overall cometary activity between millimeter and infrared ($H_2O$) observations, HCN was somewhat depleted compared with its abundance in most comets observed in the IR. However, it was consistent with that measured in the majority of comets at millimeter wavelengths, as was CS/HCN based on its value measured in ten comets. Compared with the organics-normal group, CO was severely depleted (by a factor of four or more) while $CH_3OH$ was somewhat enriched (by about 50 percent), and their abundances were consistent with those measured in the Halley-family comet 8P/Tuttle. The large abundance ratio $CH_3OH/CO$ may indicate highly efficient H-atom addition reactions on pre-cometary grain surfaces, provided that CO was not subsequently lost through nebular processing and/or locked up in more complex material such as CHON grains. Assuming measured abundances reflect processing at very low temperature ($\sim 10 - 20$ K), then compared with organics-normal comets this may indicate larger hydrogen densities in the nascent environment of C/2006 M4, or alternatively that its pre-cometary ices experienced surface chemistry over a longer period of time.

**ACKNOWLEDGMENTS**

This work was supported by the NASA Planetary Astronomy (RTOP 344-32-98) and Astrobiology (RTOP 344-53-51) Programs. Research at the Arizona Radio Observatory was supported by the NASA Astrobiology Institute under Cooperative Agreement No. CAN-02-0SS-02 issued through the Office of Space Sciences. We thank IRTF director Alan Tokunaga for granting target-of-opportunity observations on short notice, and Eric Volquardsen and Paul Sears for their expertise and assistance in conducting these difficult daytime CSHELL observations of C/2006 M4 (SWAN). The NASA-IRTF is operated by the University of Hawaii under Cooperative Agreement NCC 5-538 with the NASA-OSS Planetary Astronomy Program. The authors acknowledge the very significant cultural role and reverence that the summit of Mauna Kea has always had within the indigenous Hawaiian community. We are most fortunate to have the opportunity to conduct observations from this mountain.

| Table 1. Log of Observations of Comet C/2006 M4 (SWAN) | | | | | | |
|---|---|---|---|---|---|---|
| 2006 UT Date [a] | $R_h$ (AU) | $\Delta$ (AU) | $\Delta$-dot (km s$^{-1}$) | $\nu_0$ [b] | $t_{int}$ (min) | Species/Band/Lines targeted |
| **CSHELL/IRTF** | | | | | | |
| Nov 7.06 | 1.078 | 1.100 | +27.0 | 2985.0 | 24 | $C_2H_6$ $\nu_7$ $^rQ_0$, $^pQ_1$; $CH_3OH$ |
| 7.11 | 1.078 | 1.101 | +27.1 | 2152.5 | 28 | CO $\nu_1$ R1, R2; $H_2O$ $\nu_3$- $\nu_2$ $1_{11}$-$1_{10}$ |
| 9.11 | 1.103 | 1.134 | +30.3 | 2152.5 | 32 | " |
| 9.15 | 1.104 | 1.135 | +30.5 | 2844.0 | 8 | $CH_3OH$ $\nu_3$ Q-branch |
| 10.04 | 1.115 | 1.150 | +31.5 | 3452.25 | 36 | $H_2O$ $2\nu_1$- $\nu_1$, $\nu_1$+ $\nu_3$- $\nu_1$ (multiple lines) [d] |
| 10.10 | 1.115 | 1.152 | +31.8 | 3041.2 | 24 | $CH_4$ $\nu_3$ R1; OH* |
| 10.14 | 1.116 | 1.152 | +31.9 | 2844.0 | 8 | $CH_3OH$ $\nu_3$ Q-branch |
| **ARO 12-m** [c] | | | | | | |
| Oct 31.10 | 0.995 | 1.018 | ------ | 88631.85 | 216 | HCN J = 1 → 0 |
| Oct 31.79 | 1.004 | 1.024 | | | 276 | |
| Nov 2.97 | 1.029 | 1.045 | | | 96 | |
| Nov 10.75 | 1.124 | 1.164 | | | 180 | |
| Nov 1.79 | 1.015 | 1.033 | ------ | 146969.03 | 312 | CS J = 3 → 2 |
| Nov 3.05 | 1.029 | 1.046 | | | 96 | |
| Nov 7.76 | 1.086 | 1.111 | | | 132 | |
| Nov 9.83 | 1.112 | 1.147 | | | 350 | |

[a] Corresponds to the mid-UT decimal date. ARO results in subsequent tables pertain to the average over all dates listed.

[b] CSHELL central frequencies are expressed in cm$^{-1}$; ARO frequencies are in MHz.

[c] Beam sizes of 70 and 43 arc-seconds (translating to $5.44 \times 10^4$ and $3.29 \times 10^4$ km at the comet) were used for the HCN and CS observations, respectively.

[d] Vibrational and rotational designations for $H_2O$ lines in this setting are given in Fig. 1E.

**Table 2. Production rates in C/2006 M4 SWAN from CSHELL observations**

| UT Date 2006 | Molecule ID | $\Gamma$ [a] | $F_{line}$ [b] $10^{-19}$ W m$^{-2}$ | g-fac [d] $10^{-7}$ s$^{-1}$ | $q_{scale}$ [e] | $Q$ [f] $10^{26}$ molec s$^{-1}$ |
|---|---|---|---|---|---|---|
| Nov 07.06 | $C_2H_6$ $^pQ_1$ | 2.64±0.14 | 19.2±1.9 | 319 | 2.63±0.20 | 7.56±0.74 (1.01) |
|  | $^rQ_0$ |  | 26.7±1.9 | 355 |  | 9.43±0.66 (1.08) |
|  | **$C_2H_6$ $^pQ_1$+$^rQ_0$** |  | **45.9±2.6** | **674** |  | **8.61±0.93 (1.21)** |
| Nov 07.11 | **$H_2O$** | 1.97±0.04 | **63.1±5.2** | **5.20** | 2.20±0.41 | **1756±145 (360)** |
|  | CO R1 |  | 13.1±4.8 | 139 |  | 13.5±4.98 (5.59) |
|  | R2 |  | 6.8±4.7 | 172 |  | 5.69±3.93 (4.08) |
|  | **CO R1+R2** |  | **19.9±6.7** | **311** |  | **8.69±3.79 (4.13)** |
| Nov 09.11 | **$H_2O$** | 1.97±0.04 | **44.7±4.4** | **5.20** | 2.12±0.38 | **1280±126 (265)** |
|  | CO R1 |  | 3.1±3.4 | 139 |  | 3.29±3.64 (3.69) |
|  | R2 |  | 10.0±3.4 | 172 |  | 8.57±2.95 (3.33) |
|  | **CO R1+R2** |  | **13.1±4.8** | **311** |  | **6.47±2.58 (2.84)** |
| Nov 09.15 | **$CH_3OH$** | 1.85±0.02 | **56.1±4.7** | **103** | 1.49±0.26 | **43.3±3.65 (8.45)** |
| Nov 10.04 | $H_2O$ (1) [c] | 5.57±0.17 | 6.7±0.89 | 0.514 | 2.50±0.36 | 1470±195 (293) |
|  | (2) [c] |  | 21.2±1.2 | 1.70 |  | 1402±76.6 (222) |
|  | (3) [c] |  | 17.3±1.1 | 1.23 |  | 1582±100 (256) |
|  | (4) [c] |  | 3.0±1.5 | 0.183 |  | 1854±919 (959) |
|  | **$H_2O$ (Sum)** |  | **48.2±2.2** | **3.63** |  | **1470±58.0 (226)** |
| Nov 10.10 | **$CH_4$ R1** | 2.45±0.22 | **14.4±1.6** | **177** | 2.89±0.28 | **12.0±1.36 (2.08)** |
|  | OH* (1) [c] |  | 4.8±0.89 | 0.537 |  | 1157±215 (263) |
|  | (2) [c] |  | 1.8±0.95 | 0.350 |  | 598±314 (324) |
|  | (3) [c] |  | 4.0±0.95 | 0.153 |  | 4035±950 (1088) |
|  | (4) [c] |  | 0.67±0.96 | 0.210 |  | 222±315 (316) |
|  | **OH* (Sum)** |  | **11.3±1.9** | **1.25** |  | **880±372 (390)** |
| Nov 10.14 | **$CH_3OH$** | 1.85±0.02 | **20.4±4.8** | **103** | 4.30±0.43 | **46.4±10.9 (13.7)** |

[a] Calibration factor [$10^{-17}$ W m$^{-2}$ (cm$^{-1}$)$^{-1}$ / (ADU s$^{-1}$)], based on observations of flux standard star BS7235.

[b] Transmittance-corrected line flux contained within a 1x3 arc-second nucleus-centered aperture. Uncertainties for individual lines represent the 1σ stochastic error (for fluxes summed from multiple lines, the larger of stochastic and standard errors is listed).

[c] Water and OH prompt emission lines are identified in Fig. 1.

[d] Fluorescence g-factor at $T_{rot}$ = 80 K. For $H_2O$ on November 10, the measured value of 77 K was used (see text).

[e] Growth factor representing the measured ratio of terminal to nucleus-centered production rates.

[f] Global production rate. The first uncertainty listed is $\sigma Q_{stoch}$ (for the combined Q from multiple lines, the larger of $\sigma Q_{stoch}$, $\sigma Q_{std}$ is listed); the values in parentheses include uncertainties in $\Gamma$ and $q_{scale}$.

**Table 3. Production rates in C/2006 M4 SWAN from ARO observations**

| Molecule | $T_{rot}$ (K) | $\int T_{R*}\,dv$ [a] (K km s$^{-1}$) | $\eta_c$ [a] | $S_{ij}\|\mu_o\|^2$ ($10^{-36}$ D) | $E_j$ (MHz) | $N_{col}$ [b] ($10^{11}$ molec cm$^{-2}$) | $Q$ [b] ($10^{26}$ molec s$^{-1}$) |
|---|---|---|---|---|---|---|---|
| HCN [c] | 60 | 0.0363 | 0.9 | 8.91 | 88632 | 4.34±1.33 | 2.58±0.264 |
|  | 80 |  |  |  |  | 5.78±1.78 | 3.43±0.352 |
|  | 100 |  |  |  |  | 7.23±2.22 | 4.33±0.444 |
| CS | 60 | 0.0385 | 0.7 | 1.65 | 294060 | 3.31±1.20 | 1.31±0.159 |
|  | 80 |  |  |  |  | 4.16±1.51 | 1.64±0.199 |
|  | 100 |  |  |  |  | 5.02±1.83 | 1.99±0.242 |

[a] $T_R*$ and $\eta_c$ are measured antenna temperature and corrected beam efficiency, respectively. The quantity $T_R$ in Eq. 3 represents the equivalent brightness temperature ($T_R = T_{R*}/\eta_c$).

[b] Errors in $N_{col}$ and Q are 1$\sigma$, and include stochastic noise and uncertainties in $T_R*$ and line width. In the Monte Carlo approximation, the following parent photo-dissociation lifetimes (s) were used: HCN, $7.7 \times 10^4 R_h^2$ (Huebner et al. 1992); CS, $1.0 \times 10^5 R_h^2$ (Jackson et al. 1982); $CS_2$, $3.70 \times 10^2 R_h^2$ (Huebner et al. 1992).

[c] Values for HCN are based on the brightest component only (F = 2 –> 1; see Fig. 2A), considering it comprises 5/9 of the total (3-component) line intensity.

**Table 4. Parent volatile abundances in C/2006 M4 and other Oort cloud comets** [a]

| | CO [b] | $CH_4$ | $C_2H_6$ | $C_2H_2$ | HCN | $CH_3OH$ |
|---|---|---|---|---|---|---|
| $T_{subl}(K)$ [c] = | 24 | 31 | 42 | 57 | 95 | 99 |
| **C/2006 M4** | | | | | | |
| $T_{rot}$ = 80 K [d] | 0.48 ± 0.15 | 0.85 ± 0.18 | 0.47 ± 0.12 | --- | --- | 3.4 ± 0.69 |
| $T_{rot}$ = 60 K [d] | 0.48 ± 0.14 | 0.70 ± 0.14 | 0.49 ± 0.12 | --- | --- | 3.4 ± 0.69 |
| $T_{rot}$ = 100 K [d] | 0.46 ± 0.14 | 0.97 ± 0.15 | 0.48 ± 0.12 | --- | --- | 3.3 ± 0.68 |
| $T_{rot}$ = 80, 60 K [e] | | | | | 0.126±0.019 [e] | |
| $T_{rot}$ = 80, 80 K [e] | | | | | 0.167±0.025 [e] | |
| 4 comets [f] | 1.8 – 15 | 0.5 – 1.5 | 0.59 ± 0.03 | 0.24 ± 0.03 | 0.26 ± 0.03 | 2.2 ± 0.2 |
| C/2001 A2 | 3.9 ± 1.1 | 1.2 ± 0.2 | 1.7 ± 0.2 | 0.5 ± 0.1 | 0.6 ± 0.1 | 3.9 ± 0.4 |
| 8P/Tuttle | 0.45 ± 0.09 [g] | 0.37 ± 0.05 [h] | 0.26 ± 0.02 [h] | < 0.04 [i] | 0.07 ± 0.01 [i] | 2.7 ± 0.30 [h] |
| 1P/Halley | 3.5 | < 1 | ~ 0.4 | ~ 0.3 | ~ 0.2 | 1.7 ± 0.4 |
| C/1999 S4 | 0.9 ± 0.3 | 0.18 ± 0.06 | 0.11 ± 0.02 | < 0.12 | 0.10 ± 0.03 | < 0.15 |

[a] Values are based on $H_2O$ = 100. Uncertainties represent 1σ, and upper limits represent 3σ. Heavy (shaded red) and light (pink) solid boxes indicate respectively severely and slightly enriched abundances, and heavy (green) and light (yellow) dashed boxes indicate severely and slightly depleted abundances compared with abundances measured for the majority of Oort cloud comets (see §4.4). Unless otherwise noted, abundances are taken from DiSanti and Mumma (2008) and Mumma et al. (2003), and the original papers referenced therein should be cited.

[b] Values for CO pertain to release solely from the nucleus (i.e., the native source).

[c] Vacuum sublimation/condensation temperatures (Yamamoto 1985; Crovisier 2007).

[d] This work. Values represent mean mixing ratios for all dates measured. Uncertainties include those in Γ and $q_{scale}$ (Table 2), except when co-measured with $H_2O$, as are CO in the 2152 cm$^{-1}$ setting and $CH_4$ in the 3041 cm$^{-1}$ setting (co-measured with OH*; Fig. 1D). Values for $CH_4$ and $CH_3OH$ from November 10 incorporate the mean Q($H_2O$) as calculated from $H_2O$ in the 3452 cm$^{-1}$ setting and from OH* in the 3041 cm$^{-1}$ setting.

[e] This work. These entries use the contemporaneous value for Q($H_2O$) estimated from the activity curve of Fig. 4. For our "most probable" abundance for HCN, we take $T_{rot}$ = 80 K for $H_2O$ and 60 K for HCN, resulting in a mixing ratio of 0.126 percent as shown (see text).

[f] Refers to the mean of values for four OC comets (C/1996 B2 Hyakutake, C/1995 O1 Hale-Bopp, C/1999 H1 Lee, and 153P/Ikeya-Zhang) having similar abundances for $C_2H_6$, $C_2H_2$, HCN, and $CH_3OH$ (for abundances in individual comets, see Mumma et al. 2003). A range of values is listed for CO and $CH_4$ because the mean dispersion in their abundances among these comets (i.e., their standard distribution about the mean, or the standard error) is more than twice the error in the mean associated with their individual uncertainties of each value (the stochastic error). For $C_2H_6$, $C_2H_2$, HCN, and $CH_3OH$, the mean and the larger of stochastic and standard errors are listed.

[g] Böhnhardt et al. (2008); [h] Mean of abundances from Böhnhardt et al. (2008) and Bonev et al. (2008); [i] Bonev et al. (2008).

**Figure 1.** Molecular emissions in excess of the continuum (black traces) in C/2006 M4, indicating targeted molecules (and vibrational band, unless otherwise noted) in each panel. Modeled line intensities at the specified $T_{rot}$ (colored traces) are also shown. Vertical ticks indicate positions of lines used in our analysis, except in panel B, which shows the frequency interval over which the Q-branch of $CH_3OH$ is integrated. Also indicated are ±1σ stochastic noise (lavender traces) and the spectral extent (micrometers, in parentheses) encompassed. The $H_2O$ line near 2151 cm$^{-1}$ has vibrational band designation $\nu_3-\nu_2$ and rotational designation $1_{11}-1_{10}$. Designations for prompt OH lines numbered in panel C: (1) $\nu_1$, P12.5 1$^+$; (2) $\nu_1$, P12.5 1$^-$; (3) $\nu_2-\nu_1$, P8.5 2$^+$; (4) $\nu_2-\nu_1$, P8.5 2$^-$. Designations for water lines numbered in panel E: (1) $\nu_1+\nu_3-\nu_1$, $2_{11}-2_{20}$; (2) blend of $2\nu_1-\nu_1$, $1_{10}-2_{21}$ and $\nu_1+\nu_3-\nu_1$, $2_{02}-3_{21}$; (3) $2\nu_1-\nu_3$, $1_{10}-1_{11}$; (4) $\nu_1+\nu_3-\nu_1$, $5_{24}-6_{25}$. Inset to panel E: Boltzmann excitation diagram, showing global production rate (with ±1σ error bars) versus rotational energy (cm$^{-1}$) for the four lines. Line-by-line Qs agree (within error) at the optimal rotational temperature (77 K, see text).

**Figure 2.** Spectra of C/2006 M4 obtained with the ARO 12-m telescope. (A) HCN, with positions marked for the three nitrogen hyperfine components (labeled 'F') of the J = 1 – 0 line. The spectral resolution is 100 kHz. (B) The CS J = 3 – 2 line with a spectral resolution of 250 kHz. Both spectra are plotted in the cometocentric velocity frame.

**Figure 3**. Global water production rates in C/2006 M4 at 80 K. Points 1 and 2 are based on the $H_2O$ line near 2151 cm$^{-1}$ (Fig. 1A). Point 3 is based on the lines in Fig. 1E. Point 4 is based on the OH prompt emission lines in Fig. 1D. (For line vibrational and rotational designations, see Fig. 1.) The weighted mean Q($H_2O$) (dashed line) and its ±1σ uncertainty (dotted lines) are also indicated. Mean water production rates (from all relevant CSHELL observations) for 60, 80, and 100 K are (in units of $10^{29}$ molecules s$^{-1}$) 1.25±0.11, 1.39±0.15, and 1.53±0.16, respectively.

**Figure 4**. Overall activity light curve for C/2006 M4 (SWAN), expressed in terms of estimated water production rate as described in Section 4.2. In addition to our $H_2O$ production rates (from Fig. 3), we show UT times for the ARO observations of HCN and their weighted mean UT date (marked by the arrow). The estimated $H_2O$ production rate at this time (2.05 x $10^{29}$ molecules s$^{-1}$) is indicated by the horizontal dashed line. The weighted mean UT date for the CS observations is also shown.

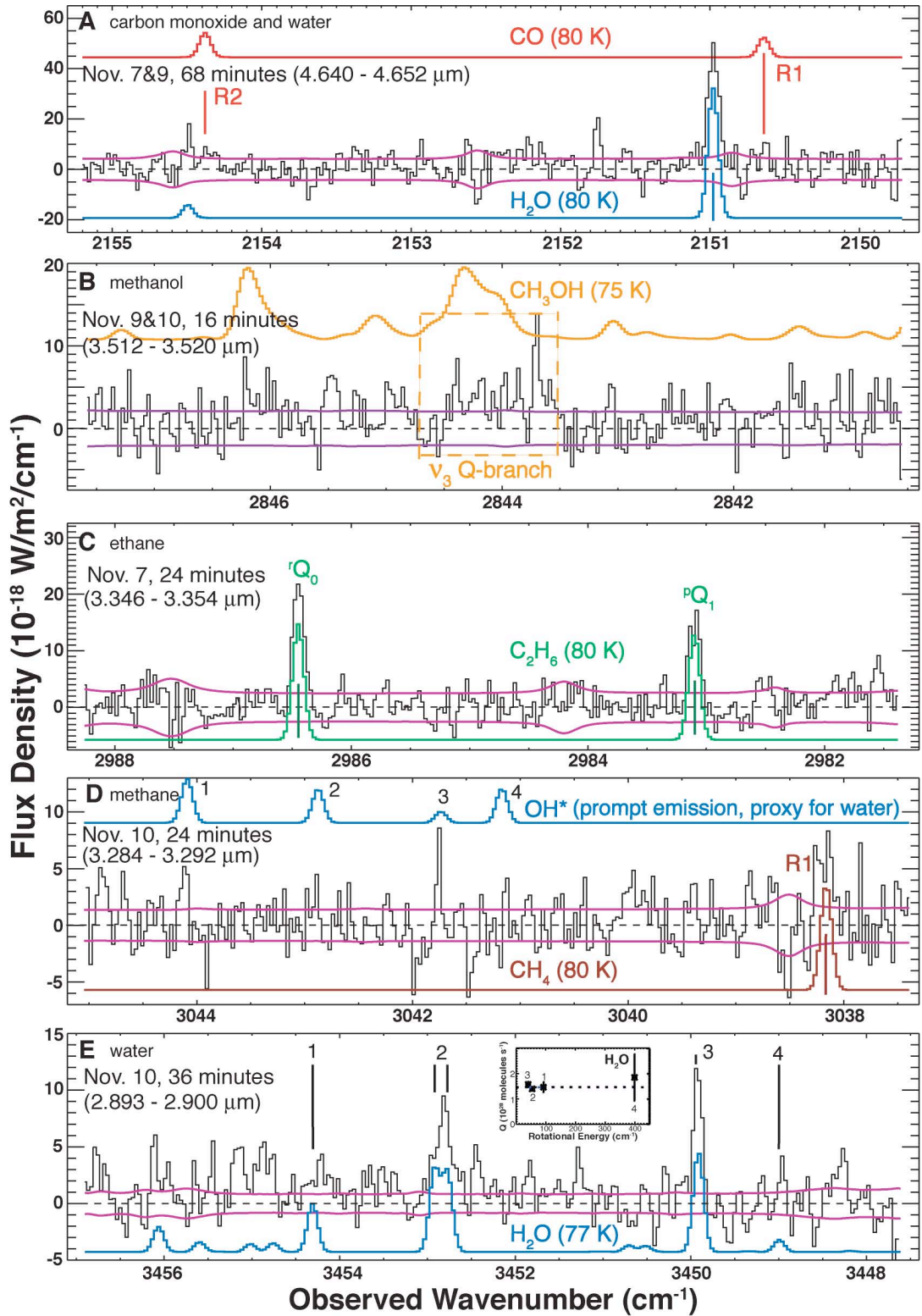

DiSanti et al. C/2006 M4 SWAN, Fig. 1

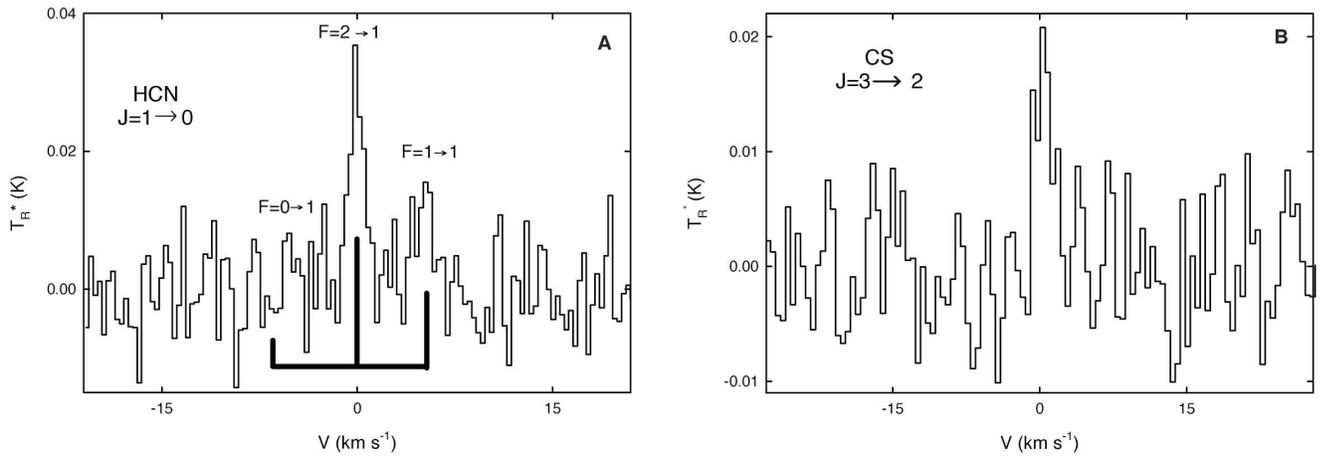

DiSanti et al. C/2006 M4 SWAN, Fig. 2

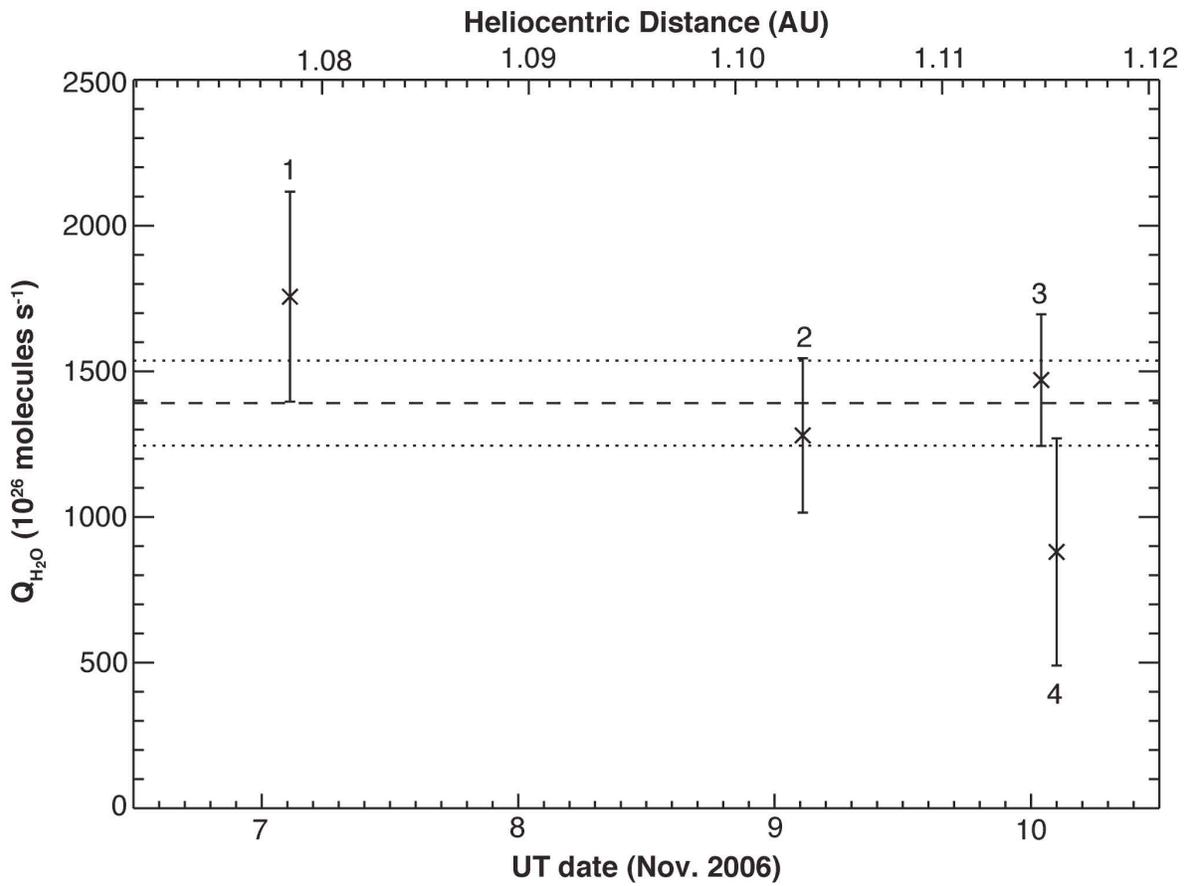

DiSanti et al. C/2006 M4 SWAN, Fig. 3

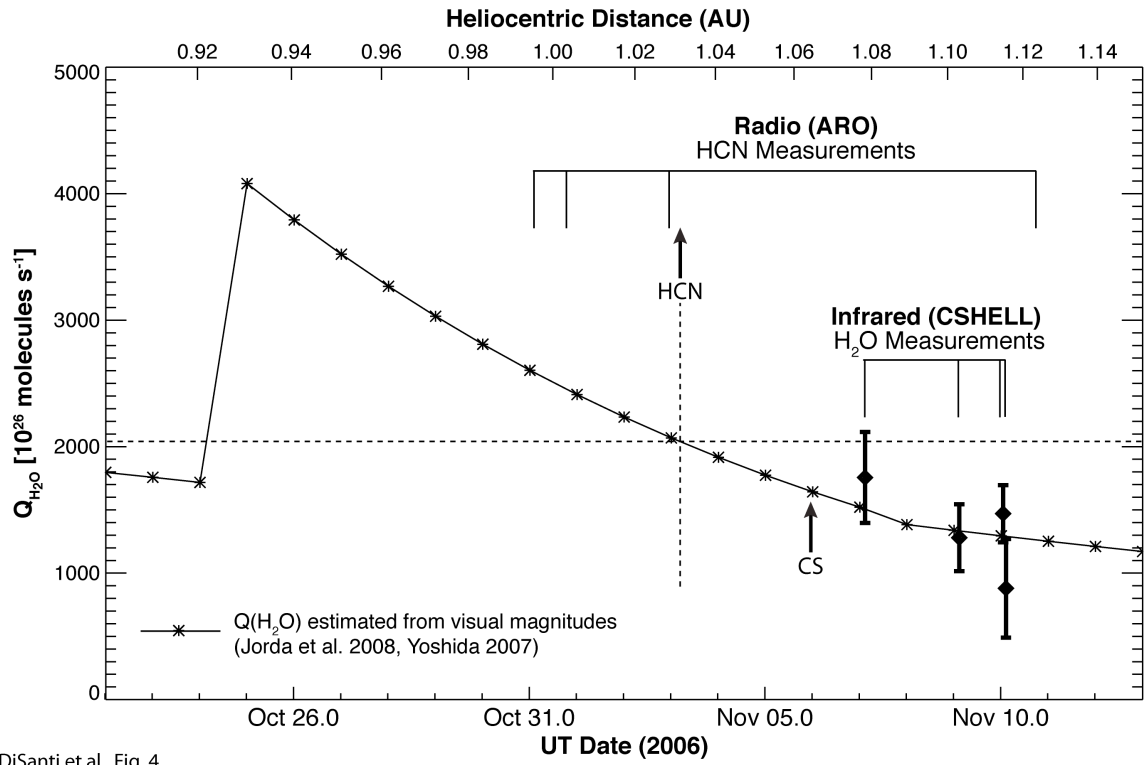

DiSanti et al., Fig. 4